\documentclass[%
 reprint,
superscriptaddress,
 nofootinbib,
 amsmath,amssymb,
 aps,
 floatfix,
 pre
]{revtex4-2}

\usepackage[dvipsnames]{xcolor}
\usepackage{graphicx}
\usepackage{dcolumn}
\usepackage{bm}
\usepackage{comment}
\usepackage{hyperref}

\usepackage{todonotes}
\setuptodonotes{size=\footnotesize}

\usepackage{pgfplots}
\pgfplotsset{compat=newest}

\usepackage{tikz}
\usetikzlibrary{arrows, arrows.meta, backgrounds, external}

\usepackage{siunitx}
\DeclareRobustCommand{\mean}[1]{\ensuremath{ { \langle #1 \rangle} }}
\DeclareRobustCommand{\vect}[1]{\ensuremath{\boldsymbol{#1}}}
\newcommand{\dd}{\ensuremath{\,\mathrm{d}}} 

\begin{document}

\preprint{APS/123-QED}

\title{Anomalous Diffusion in the Square Soft Lorentz Gas}

\author{Esko Toivonen}
 \email{esko.toivonen@tuni.fi}
\affiliation{Computational Physics Laboratory, Tampere University, P.O. Box 600, FI-33014 Tampere, Finland}

\author{Joni Kaipainen}%

\affiliation{Computational Physics Laboratory, Tampere University, P.O. Box 600, FI-33014 Tampere, Finland}
\affiliation{Integrated Computational Materials Engineering group, VTT Technical Research Centre of Finland Ltd, FI-02044, Espoo, Finland}

\author{Matti Molkkari}
\affiliation{Computational Physics Laboratory, Tampere University, P.O. Box 600, FI-33014 Tampere, Finland}

\author{Joonas Keski-Rahkonen}%
\affiliation{Computational Physics Laboratory, Tampere University, P.O. Box 600, FI-33014 Tampere, Finland}
\affiliation{Department of Physics, Harvard University, Cambridge, Massachusetts 02138, USA
}%
\author{Rainer Klages}%
\affiliation{Centre for Complex Systems, School of Mathematical Sciences, Queen Mary University of London, Mile End Road, London E1 4NS, United Kingdom}
\affiliation{London Mathematical Laboratory, 8 Margravine Gardens, London W6 8RH, UK}
\author{Esa Räsänen}%
 \email{esa.rasanen@tuni.fi}
\affiliation{Computational Physics Laboratory, Tampere University, P.O. Box 600, FI-33014 Tampere, Finland}

\date{\today}

\begin{abstract}
We demonstrate and analyze anomalous diffusion properties of point-like particles in a two-dimensional system with circular scatterers arranged in a square lattice and governed by smooth potentials, referred to as the square soft Lorentz gas. Our numerical simulations reveal a rich interplay of normal and anomalous diffusion depending on the system parameters. To describe diffusion in normal regimes, we develop a unit cell hopping model that, in the single-hop limit, recovers the Machta-Zwanzig approximation and converges toward the numerical diffusion coefficient as the number of hops increases. Anomalous diffusion is characterized by quasiballistic orbits forming Kolmogorov-Arnold-Moser islands in phase space, alongside a complex tongue structure in parameter space defined by the interscatterer distance and potential softness. The distributions of the particle displacement vector show notable similarities to both analytical and numerical results for a hard-wall square Lorentz gas, exhibiting Gaussian behavior in normal diffusion and long tails due to quasiballistic orbits in anomalous regimes. 
Our work thus provides a catalog of key dynamical system properties that characterize the intricate changes in diffusion when transitioning from hard billiards to soft potentials.
\end{abstract}

\maketitle

\section{Introduction}\label{sec:level1}

Two-dimensional (2D) materials and electronic systems have emerged as a frontier in solid-state physics, offering promising applications in electronics, such as transistors and memory devices~\cite{zeng2021}. The physics of these systems, particularly when far from the thermodynamic limit -- referred to as {\em small systems} \cite{smallsys1,smallsys2} with only a few relevant degrees of freedom -- is characterized by nonequilibrium transport and nonlinear dynamics that can result in chaotic behavior. This gives rise to a diverse range of macroscopic phenomena, such as branched flow~\cite{Heller2021,Daza2021, graf2024chaos}, or anomalous diffusion, where the mean squared displacement (MSD) scales nonlinearly with time~\cite{KRS08,MJJCB14,ZDK15,Oliveira2019}.

The Lorentz gas is a simple yet powerful model for examining the diffusion properties of 2D systems~\cite{Sza00,Klages_book,Dettmann_2014}. It consists of a point particle that scatters elastically with fixed hard spheres, which are distributed either randomly or periodically in space. Originally put forward in 1905~\cite{lorentzgas} to replicate Drude’s theory of electrical conductivity, the Lorentz gas has been instrumental in mathematical physics. It has been employed, for example, in proving Ohm’s law~\cite{ohm2}, studying Lyapunov exponents~\cite{Do99} and fractal attractors~\cite{HoB99}, and formulating the chaotic scattering theory of transport~\cite{Gaspard1998}.

Conventionally, scatterers in the Lorentz gas are modeled as circular hard-wall potentials. However, various other potentials have also been explored, including trigonometric functions~\cite{GZR88,lorentzcos1}, Lennard-Jones potentials~\cite{lorentzlennard,YaZh10}, Coulomb potentials \cite{Nob95}, WCA potentials \cite{PeFr19}, and hard-wall disks with low potential regions~\cite{lorentzlowpot}
Recent work by some of the present authors~\cite{PRL} has focused on a {\em soft} Lorentz gas, which smoothly extrapolates from hard to soft walls under the variation of a potential parameter. This is accomplished by replacing the hard-wall disks with Fermi-type potential profiles arranged in a triangular lattice. 
This soft potential is particularly relevant as a model for various 2D electronic systems, such as artificial graphene~\cite{Gomes2012,Rasanen2012,Paavilainen2016,Polini2013}, where the confining potential is smooth. It was shown that the triangular soft Lorentz gas exhibits both normal and anomalous diffusion, with extreme sensitivity to the model parameters~\cite{PRL}. This complex interplay between spatially localized periodic orbits and quasiballistic trajectories is characterized by tongue structures as functions of the system parameters. 
Quasiballistic trajectories are defined as trajectories that are regular and propagate infinitely towards one direction.
Similar to these findings, recent studies on branched flow~\cite{Daza2021, graf2024chaos} have demonstrated that at high energies, significantly exceeding the amplitude of the scatterers, the dynamics remain strongly dependent on the potential shape. Notably, soft Fermi-type potentials are able to induce branched flow even in a periodic lattice~\cite{Daza2021}.

In this work, we focus on the soft Lorentz gas arranged in a square lattice. This system is experimentally realizable in various antidot superlattices, fabricated from materials such as GaAs/AlGaAs heterostructures \cite{LKP91,Weis91,shi2015antidot} or graphene~\cite{yagi2015graphenetri,sandner2015graphenesqu}.
In contrast to the triangular configuration previously studied in Ref. \cite{PRL}, the square lattice has an \emph{infinite horizon} for \emph{all} lattice spacings, meaning there are singular channels in the billard along which a particle can travel in a straight line to infinity \cite{machta_zwanzig,klages2000}. For soft potentials the speed fluctuates along these paths due to variations of the potential, and the corresponding trajectories are related to the emergence of regular structures in the phase space \cite{GZR88,turaev1998islands,RKT99}.
The existence of an infinite horizon within the lattice complicates the exercise of traditional approximations, as for hard wall billiards it implies that normal diffusion does not exist \cite{Sza00,Klages_book,Dettmann_2014}.

Furthermore, the conventional Machta-Zwanzig (MZ) approximation assumes that hops between unit cells are uncorrelated~\cite{machta_zwanzig}. This assumption as well as a well-defined hard wall normal diffusive limit do not hold in a generic infinite-horizon lattice, calling for the development of more refined approximations.
While a Taylor-Green-Kubo-based expansion has been proposed for triangular lattices~\cite{klages2000,Klages2002}, as well as other approaches based on stochastic theory \cite{GilSan09,GNS11,CGLS14,CGLS15}, we here introduce an alternative approximation that accounts for correlations between unit cell hops. The established model is then compared against the widely applied MZ approximation, as well as the numerical results for the diffusion coefficient.

In particular, we find several common features in the anomalous diffusion of square and triangular systems. Similarly as in the case of the triangular arrangement, we analyze the tongue structure in detail, including their relation with the existence of Kolmogorov-Arnold-Moser (KAM) islands, and discover complex fine structure in the density of quasiballistic orbits. Furthermore, we emphasize the weak-softness limit of the system, and find a qualitative agreement with the properties of the conventional (hard-wall) Lorentz gas. However, the extreme sensitivity of the system to potential parameters is also demonstrated in this limit.

The paper is organized as follows. In Sec.~\ref{sec:model}, we introduce the model and the numerical tools used for its analysis. Section~\ref{sec:theory} presents the formulation of a hopping model approximation for the diffusion coefficient. Our numerical results are detailed in Sec.~\ref{sec:results}, where we first discuss the diffusion coefficient, followed by an analysis of anomalous diffusion regimes, quasiballistic trajectories, phase-space structures, and tongue-like features in the parameter space. We conclude with a discussion of particle displacement distributions. The paper is summarized in Sec.~\ref{sec:summary}.

\section{Model and methodology}\label{sec:model}

\begin{figure}[ht]
\centering
\includegraphics{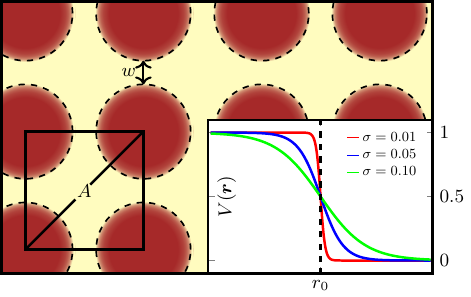}
\caption{Visualization of a square soft Lorentz gas. The periodic unit cell is labeled as region $A$, and the diagonal inside the unit cell corresponds to the Poincaré surface of section used later in the article. The effective gap between the scatterers is denoted by $w$. The inset shows the cross section of the Fermi potential for a few values of the softening parameter $\sigma$.}
\label{fig:geometry}
\end{figure}

Our soft Lorentz gas is defined as a square lattice of circular scatterers modeled as a smooth Fermi potential:
\begin{equation}
    V(\bm{r})=\frac{V_0}{1+\exp\left(\frac{|\bm{r}|-r_0}{\sigma}\right)},
\end{equation}
where $r_0$ is the effective radius of each scatterer and
$\sigma$ is a softening parameter. For simplicity, we can choose the energy scale of the system in such a way that the potential amplitude of the scatterers is $V_0 = 1$. The lattice, unit cell and the potential profile of the scatterers are illustrated in Fig.~\ref{fig:geometry}.
In Hartree atomic units ($r_0=m=1$), the total energy of the particle is set to be $E=1/2$.
We expect the results obtained with this energy choice to be somewhat representative for all energies for which particles can escape the unit cell but, on the other hand, cannot cross the potential maxima ($V_0 = 1$). Above~$E = 1$, the energy-dependent diffusion exhibits partially somewhat different, novel features, as has been explored in detail for the triangular lattice~\cite{gilgallegos2019}.
The periodic unit cell has a side length of  $2r_0+w$, where $w$ is the nominal gap between the scatterers (measured at a half-maximum of the Fermi potential). Below we consider $w$ and $\sigma$ as the two main controlled parameters. The limit
$\sigma\rightarrow 0$ corresponds to the conventional hard-wall Lorentz gas, whereas when~$\sigma$ is large, the system approaches the integrable cosine potential.

Our main quantity of interest is the 2D diffusion coefficient $D$ given by
\begin{equation}\label{eq:einstein}
    D=\lim_{t \to \infty}\frac{\langle(\bm{r}(t)-\bm{r}(0))^2\rangle}{4t},
\end{equation}
where $\langle(\bm{r}(t)-\bm{r}(0))^2\rangle$ is the MSD for position $\bm{r}(t)$ of a particle at time $t$. Here the angular brackets denote a configurational ensemble average. In the case of normal diffusion, the MSD grows linearly in time, i.e., ${\rm MSD}\propto t$.
Superdiffusion and subdiffusion are characterized by ${\rm MSD}\propto t^{\alpha}$ with $\alpha>1$ and $\alpha<1$,
respectively, and ballistic motion corresponds to $\alpha=2$.
We point out that Eq.~\eqref{eq:einstein} is only valid for {\em normal}, i.e., non-anomalous diffusion. In the following, however, we apply this definition numerically to various scenarios to identify anomalous diffusion, which will be analyzed separately.

Determining $D$ using Eq.~(\ref{eq:einstein}) requires computing the MSD with a sufficiently large ensemble. In the simulations, we employ the Bill2D software package~\cite{bill2d} for different values of $w$ and $\sigma$. To ensure the accuracy of the simulations, we use the 6th-order symplectic integrator for the equations of motion~\cite{theintegrator}.
This integrator produces the most accurate conservation of energy within the parameter space of our simulations.
See the Appendix for details on the evaluation of different propagation algorithms and their timesteps.
A 3-by-3 unit cell grid is formed to accurately compute the potential values for a particle in the centermost unit cell.
The initial positions of the particles are uniformly randomized in the energetically allowed coordinate space in the unit cell, and the initial speeds of the particles are determined based on the constant energy condition of~$E=1/2$ while the initial propagation angles are randomized.
This procedure is utilized in Sec.~\ref{sec:results} unless otherwise specified.

\section{Theoretical predictions}\label{sec:theory}

In previous studies~\cite{Klages_book,machta_zwanzig,klages2000,Dettmann_2014,Klages2002,GilSan09,GNS11,CGLS14,CGLS15}, numerous analytical approximations have been brought into play to compute the diffusion coefficient $D$ in the conventional Lorentz gas. In many of these predictions, random walks together with diffusive processes are reviewed to obtain $D$ as a function of the density of the Lorentz gas scatterers.

Following the reasoning behind the MZ approximation~\cite{machta_zwanzig}, we associate the diffusion with random hopping between adjacent unit cells.
By the phase space argument, the mean residence time $\tau$ within a unit cell is approximately
\begin{equation}\label{eq:phaseapprox}
    \tau \approx \frac{\Omega}{\omega} \text{,}
\end{equation}
where $\Omega$ is the total allowed phase space volume within the unit cell -- henceforth called a ``trap'' -- and $\omega$ is the outwards phase space flux from the trap.
Thus, integrating over both velocity and position spaces yields
\begin{align}
    \Omega &= 2\pi \int_{V(\vect{r}) \leq E} v(\vect{r}) \dd{\vect{r}} = 2\pi A_{\text{trap}} \mean{v(\vect{r})}_{\text{trap}}, 
\end{align}
where $v(\vect{r}) = \sqrt{2\left(E - V(\vect{r})\right)}$ is the magnitude of the velocity obtained from the conservation of energy $E$, $A_{\text{trap}}$ is the area of the trap, and $\mean{v(\vect{r})}_{\text{trap}}$ is the average magnitude of the velocity within the trap.
Utilizing the symmetry of the four exits from the trap, the total outwards phase space flux is given by
\begin{align}
    \omega &= 4 \int_{V(x) \leq E} \int_{-\frac{1}{2}\pi}^{\frac{1}{2}\pi} v(x)^2 \cos\theta \dd\theta \dd x \\
    &= 8 l_{\text{exit}} \mean{v(\vect{r})^2}_{\text{exit}} \text{,}
\end{align}
where $V(x)$ and $v(x)$ are evaluated at $\vect{r}=(x, 0)$, $l_{\text{exit}}$ is the length of the exit, and $\mean{v(\vect{r})^2}_{\text{exit}}$ is the mean squared magnitude of the velocity within the exit. Note that the integrand is the velocity space density multiplied by the dot product of the velocity vector and the outwards pointing unit normal vector of the exit, yielding the outwards phase space flux at the given point $x$ towards $\theta$, and the total flux when integrated.

By assuming that the hopping between the traps is uncorrelated, the diffusion coefficient is then given by 
\begin{equation}
    D_{\text{MZ,soft}} = \frac{l^2}{4\tau} = \frac{\left(2r_0 + w\right)^2 l_{\text{exit}} \mean{v(\vect{r})^2}_{\text{exit}}}{\pi A_{\text{trap}} \mean{v(\vect{r})}_{\text{trap}}} \label{eq:soft potential diffusion coefficient}
\end{equation}
in terms of the lattice spacing $l = 2r_0 + w$. In general, there is no analytical solution for $A_{\text{trap}}$, $l_{\text{exit}}$, or the averages, but it is straightforward to compute them numerically.
We include this approach as our first-level approximation in Fig.~\ref{fig:dcoef}(a), as we expect that the uncorrelatedness assumption is violated due to the presence of an infinite horizon in the square lattice.

A more precise formulation for the diffusion coefficient can therefore be obtained by considering unit cell hopping sequences of length $n$ and calculating the diffusion coefficient using the path lengths and durations, weighted by the probability of the hopping paths.
Here, a hop is defined as exiting a unit cell in a certain (absolute) direction.
We define~$\mathcal{H} = \left\{\leftarrow, \uparrow, \rightarrow, \downarrow \right\}$ as the alphabet of possible hopping directions, and~$\Delta_i \in \mathcal{H}$ as the direction of the $i$th hop.
Next, we determine the displacement vector covered by a certain path~$\Delta_1\Delta_2\cdots\Delta_n$ of length~$n$ as 
\begin{equation}
    \vect{R}(\Delta_1 \cdots \Delta_n) = \sum_{i = 1}^n \vect{\rho}_{\Delta_i}
\end{equation}
with
\begin{align*}
    \vect{\rho}_{\leftarrow} &= \begin{bmatrix}-1 \\ 0\end{bmatrix}
    &\vect{\rho}_{\uparrow} &= \begin{bmatrix}0 \\ 1\end{bmatrix}
    &\vect{\rho}_{\rightarrow} &= \begin{bmatrix}1 \\ 0\end{bmatrix}
    &\vect{\rho}_{\downarrow} &= \begin{bmatrix}0 \\ -1\end{bmatrix} \text{.}
\end{align*}
Subsequently, the squared distance traveled by the path is~$\vect{R}^2(\Delta_1 \cdots \Delta_n)$, where squaring implies the dot product.
By computing the MSD over all paths of length~$n$ by weighting the paths by their probabilities, we obtain our higher-tier approximation for the time-dependent diffusion coefficient
\begin{equation}
    D^{n\text{-hop}} = \frac{l^2}{4n\tau} \sum_{\substack{\text{Permutations}\\\Delta_1 \cdots \Delta_n}}  p(\Delta_1 \cdots \Delta_n) \vect{R}^2(\Delta_1 \cdots \Delta_n) \text{,}
\end{equation}
where the residence time of any path is taken from the phase space approximation according Eq.~\ref{eq:phaseapprox}.

The residence time~$\tau$ is not constant for different paths due to the infinite horizon.
In particular, hopping paths passing straight through several unit cells are expected to take less time than the other paths.
Hence, we take the path-dependent residence times into account, which comes with the following implicit assumption: the MSD of any individual $n$-hop path scales linearly with time during the time range of the $n$ hops. Hence, we obtain
\begin{equation}
    \label{eq:unit cell hopping dcoeff}
    D^{n\text{-hop}}_\text{MZ} = \frac{l^2}{4} \sum_{\substack{\text{Permutations}\\\Delta_1 \cdots \Delta_n}} \frac{ p(\Delta_1 \cdots \Delta_n) \vect{R}^2(\Delta_1 \cdots \Delta_n) }{\tau(\Delta_1 \cdots \Delta_n)} \text{,}
\end{equation}
where~$\tau(\Delta_1 \cdots \Delta_n)$ is the time taken by a certain path and~$p(\Delta_1 \cdots \Delta_n)$ is the probability of the path.
The sum runs over all unique hopping paths of length~$n$, consisting of hoppings as defined by alphabet~$\mathcal{H}$. In general, we denote the diffusion coefficients $D_{\text{MZ}}^{n\text{-hop}}$ utilizing this improved approximation.
Since there are no known analytical results for the probabilities or the exact residence times, we are required to estimate them based on simulation data.

In the special case of $n=1$, our $n$-hop approximation of the diffusion coefficient renders to the standard result of the MZ approximation. To verify this fact, we expand the sum, yielding
\begin{align}
    D_{\text{MZ}}^{1\text{-hop}} = \frac{l^2}{4} \Biggl(
       &\frac{ p(\uparrow   ) \vect{R}^2(\uparrow   ) }{ \tau(\uparrow   ) } + 
        \frac{ p(\rightarrow) \vect{R}^2(\rightarrow) }{ \tau(\rightarrow) } + \nonumber\\
       &\frac{ p(\downarrow ) \vect{R}^2(\downarrow ) }{ \tau(\downarrow ) } + 
        \frac{ p(\leftarrow ) \vect{R}^2(\leftarrow ) }{ \tau(\leftarrow ) }
    \Biggr).
\end{align}
Since the residence times~$\tau$ are equal, the squared distances are unities, and each probability is~$\frac{1}{4}$, this pathway ergo leads to the anticipated MZ expression:
\begin{equation} 
    D_{\text{MZ}}^{1\text{-hop}} = \frac{l^2}{4} \Biggl(\frac{1}{4}\frac{1}{\tau} + \frac{1}{4}\frac{1}{\tau} + \frac{1}{4}\frac{1}{\tau} + \frac{1}{4}\frac{1}{\tau}\Biggr) = \frac{l^2}{4\tau}.
\end{equation}

To the best of our knowledge, this approach has not previously been applied to the calculation of the diffusion coefficient in a periodic system. We want to emphasize that this way is \emph{not} a Taylor-Green-Kubo-based expansion, such as in~Ref.~\cite{Klages2002}.
Both approaches define sequences (indexed by the hopping path length $n$) of time-dependent diffusion coefficients that converge -- if the assumptions behind the approximations hold -- to the exact diffusion coefficient, based on encoding the paths of diffusing particles in terms of symbol sequences which map the original dynamics onto correlated random walks on a lattice.
However, the present approach does so with respect to a properly weighted MSD, where the weight is taken as the probability of a certain hopping path of length~$n$ occurring at a corresponding average time $\tau(\Delta_1 \cdots \Delta_n)$.
Instead, the Taylor-Green-Kubo approach evaluates the velocity autocorrelation function along the very same hopping path of length $n$ but with respect to multiples $n$ of the mean residence time $\tau$.
Moreover, we do not consider lattice vectors but unit cell hoppings.

\section{Numerical results}\label{sec:results}

\subsection{Diffusion coefficient}

\begin{figure}[ht]
    \includegraphics{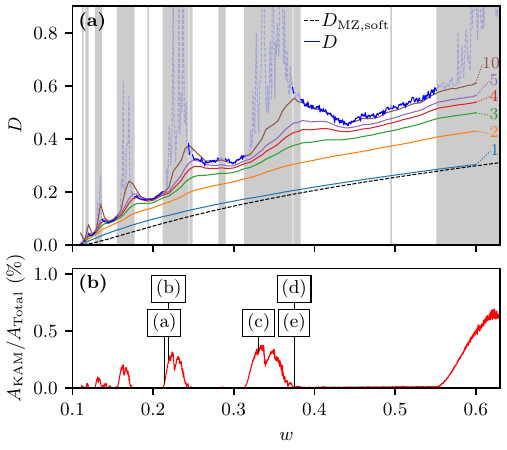}
    \caption{
        (a) Diffusion coefficient $D$ as a function of the distance between the scatterers $w$ in a square soft Lorentz gas with softness parameter $\sigma=\num{0.05}$. The numerical results (blue solid line) are compared to the Machta-Zwanzig approximation of Eq.~(\ref{eq:soft potential diffusion coefficient}) (dashed line) and the hopping approximation of Eq.~(\ref{eq:unit cell hopping dcoeff}) (numbered solid lines). 
        The grey areas correspond to quasiballistic regions (see text). (b) Corresponding proportion of KAM islands of the Poincaré surface of section. The labels (a-e) refer to the KAM islands and trajectories in Fig.~\ref{fig:kam}.
    }
    \label{fig:dcoef}
\end{figure}

We begin by numerically investigating the diffusion coefficient $D$ as a function of the interscatterer distance $w$ with a fixed softness parameter $\sigma=\num{0.05}$. The results are collectively presented in Fig.~\ref{fig:dcoef}(a).
The black dashed line denotes the MZ approximation according to Eq.~\eqref{eq:soft potential diffusion coefficient}. The numbered solid lines label the results of the estimations given by our upgraded hopping approximation defined in Eq.~\eqref{eq:unit cell hopping dcoeff} with different hopping path lengths. Finally, the blue solid line shows our numerical results for $D$ computed from $N=\num{20 000}$ trajectories for each~$w$. Here~$D$ is computed from $\num{5000}$ to $\num{10000}$ time units.

In the gray regions, the diffusion constant $D$ in the sense of Eq.~\ref{eq:einstein} is ill-defined due to the existence of quasiballistic trajectories (at least one in the ensemble). They are identified by a criterion that the squared distance metric is monotonically increasing for at least $99\,\%$ of the simulation time. The dashed blue line in the gray regions shows the results computed without these quasiballistic trajectories. 

We find that the hopping model converges towards the numerically computed diffusion coefficient; whereas the MZ approximation underestimates the actual diffusion coefficient roughly by a factor of two. With path length~$n=\num{10}$, the improved hopping estimation is already close to the numerical result in the regions where the diffusion is predominantly normal. In those regions the behavior of numerically computed $D(w)$ is complex with a clear fine structure (see below for details) and qualitatively similar to that of a triangular soft Lorentz gas~\cite{PRL}.

Let us next briefly assess the limits of small or large gap $w$. We find in 
Fig.~\ref{fig:dcoef}(a) that below a critical value $w_c$, which depend on the softness $\sigma$, no diffusion exists as the particles stay energetically trapped inside a single unit cell. We can estimate this critical gap-softness ratio to be
\begin{equation}\label{eq:critical_value}
    \frac{w}{\sigma} \lesssim 2 \ln \left( \frac{2V_0}{E} - 1\right)
\end{equation}
by assuming that the two nearest Fermi scatterers of height of $V_0$ are the only culprits blocking the escape channel off the unit cell for the particle with fixed energy $E$. For $\sigma = \num{0.05}$ (with $E =1/2$ and $V_0 = 1$ as chosen previously), the estimation yields the critical width $w_c = 2\sigma\ln(3) \approx \num[round-precision=6,round-mode=places]{0.1098612289}$, which coincides extremely well with the numerical simulations. Moreover, this estimation concurs with the later reported no-diffusion area in Fig.~\ref{fig:tongue}(a). On the other hand, at $w \approx \num{0.55}$ we find an increasing number of quasiballistic trajectories. We computed numerically that the rightmost gray region in Fig.~\ref{fig:dcoef}(a) continues to at least $w=\num{4.0}$.

\subsection{Phase space structure and KAM islands}\label{sec:kam}

The quasiballistic regions in Fig.~\ref{fig:dcoef}(a) can be understood and further analyzed in the terms of the rise and fall of KAM islands in the phase space. We define the Poincaré surface of section (PSOS) as the main diagonal of the unit cell, as visualized in Fig.~\ref{fig:geometry}. For each crossing, we obtain variables~$(s, v_\parallel)$ denoting the distance along this diagonal and the velocity parallel to the diagonal at crossing time, respectively. 
Figure~\ref{fig:dcoef}(b) shows the proportion of KAM islands in the surface of section as a function of $w$. The existence of KAM islands qualitatively agrees with the quasiballistic regions in Fig.~\ref{fig:dcoef}(a). 
However, we want to point out that the KAM islands are generally tiny, covering under one percent of the entire energetically allowed phase space area in the PSOS. 
This consequently complicates the detection of these islands, requiring extensive phase space sampling.
Furthermore, it is important to notice that some of the KAM islands correspond to localized periodic orbits (see below). At $w \approx \num{0.55}$, the observed KAM islands appear as persistent, explaining the continuous gray region in Fig.~\ref{fig:dcoef}(a).

\begin{figure}[ht]
    \includegraphics{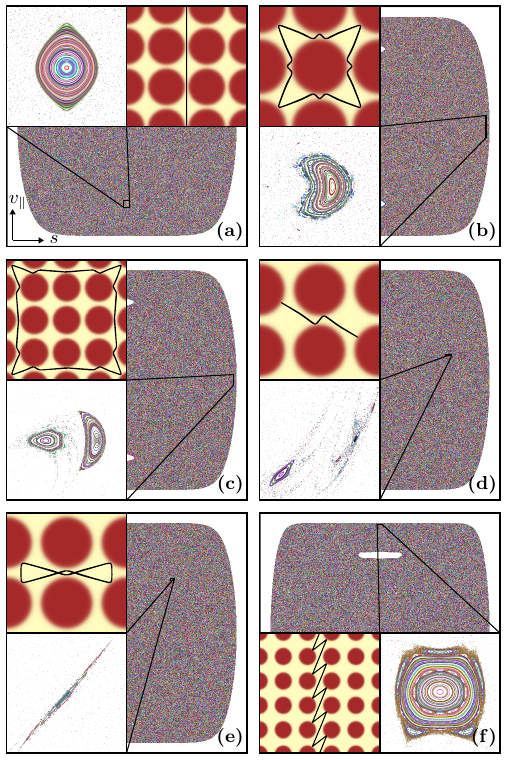}
    \caption{
        Examples of KAM islands and corresponding trajectories. The white areas inside the Poincaré surfaces of section are KAM islands, but no trajectories have been sampled there. The smoothness of the potential is fixed to $\sigma = 0.05$, but the interscatterer distance $w$ varies as follows: (a) $w = \num{0.2140}$,
        (b) $w = \num{0.2195}$
        (c) $w = \num{0.3310}$
        (d--e) $w = \num{0.3748}$
        (f) $w = \num{0.8535}$
    }
    \label{fig:kam}
\end{figure}

The appearance of KAM islands in our soft Lorentz gas is expected, as it has been proven that any perturbation of a scattering billiard system to a smooth one may cause KAM islands to form in the vicinity of the singular periodic orbits of the original billiard.~\cite{turaev1998islands,RKT99} The corresponding hard-wall square Lorentz gas has ballistic orbits along the main coordinate axes, and these orbits have a measure of zero in the phase space. Softening the potential transform these orbits into islands of stability. As $w$ increases, these islands become wider as well as thinner, suggesting that at the limit~$w\to\infty$ we might recover the hard-wall case, i.e., a straight line along the PSOS.

In Fig.~\ref{fig:kam}, we present some examples of KAM islands in the regimes labeled (a--e) in Fig.~\ref{fig:dcoef}(b), with an increasing order of $w$. The main figures show the entire allowed phase space, and the insets show a zoomed PSOS exhibiting the KAM island, together with a corresponding trajectory in that island. This analysis confirms and illustrates the previous statement that the KAM islands are typically minuscule, as already demonstrated by the proportional areas in Fig.~\ref{fig:dcoef}(b).

Figure~\ref{fig:kam}(a) shows a typical PSOS for nearly straight quasiballistic trajectories. The minor oscillation in the example trajectory is not visible without extreme distortion and zoom in the figure. The trajectories corresponding to the center of the KAM island have the lowest amplitude in the oscillations.
If $w$ is increased, the KAM island gradually becomes larger, deforms, bifurcates into two islands, and finally disappears. This behavior is illustrated in the video included in the Supplemental Material~\cite{SM}. The top-right panel shows the complete PSOS, with a magnified section highlighted in the left panel, while the bottom-right panel presents the values of~$D$ and the area of the KAM islands within the PSOS.
The black vertical line indicates the current value of~$w$ displayed in the PSOS panels.
The white strip at $w\sim\num{0.24}$ is due to the fact that no quasiballistic trajectories have been sampled at that value of~$w$.

Next, Figs.~\ref{fig:kam}(b--e) show some specimens of localized periodic orbits.
The trajectories in Figs.~\ref{fig:kam}(b--c) periodically circulate around one and nine scatterers, respectively. We note that even though only one KAM island is shown in detail, the trajectory forms several small islands in the PSOS. We further conjecture that similar circulating trajectories around a square arrangement of scatterers exist, but they are rare with minute KAM islands.

Figures~\ref{fig:kam}(d--e) show two localized periodic orbits, which 
notably appear with the same parameter values ($\sigma=\num{0.05}$, $w=\num{0.3748}$). For the needle-like trajectory in Fig.~\ref{fig:kam}(d) all the rotations and mirror versions are present, and the following KAM island structure is rather intricate.
The bowtie-like trajectory in Fig.~\ref{fig:kam}(e) forms a diagonal KAM island in the PSOS.

As the final example, Fig.~\ref{fig:kam}(f) shows a zigzag-like quasiballistic orbit. Here the Lorentz gas is already relatively sparse with $w = \num{0.8535}$, and the KAM island formed by this trajectory is diminutive.

\subsection{Tongue structures}

\begin{figure}[ht]
    \includegraphics{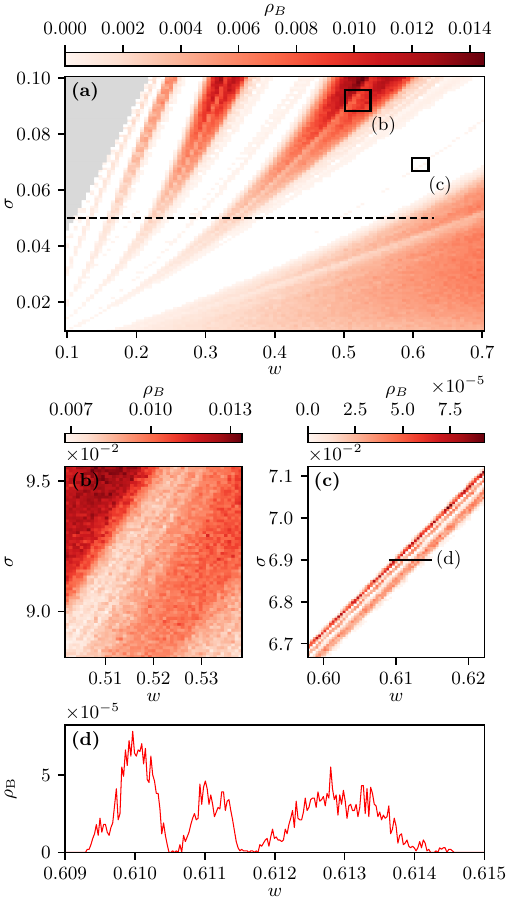}
    \caption{
        (a) Density of quasiballistic trajectories over the parameter space of the square soft Lorentz gas, characterized by a tongue-like structure. The grey area in top left corresponds to the parameter space where the particles cannot escape from the unit cell, and the dashed horizontal line refers to Fig.~\ref{fig:dcoef}.
        (b-c) Zoomed sections of (a) corresponding to high and low densities of quasiballistic trajectories, respectively. Note the different colorbars in (b) and (c).
        (d) Zoomed strip of (c), illustrating the irregular cross-section of the tongue. In (c) and (d), the ensemble size has been increased to~$\num[print-unity-mantissa=false]{e6}$.
    }
    \label{fig:tongue}
\end{figure}

As the next step, we explore the 2D parameter space~$(w, \sigma)$ to identify and characterize the regimes that display anomalous diffusion. 
Figure~\ref{fig:tongue}(a) shows the relative proportion, i.e., the \emph{density} of quasiballistic trajectories,~$\rho_\mathrm{B}$.
Each pixel in Fig.~\ref{fig:tongue}(a) is covered by $N = \num{100000}$ simulations, with a simulation time of~$t = \num{1000}$. This sampling size ensures sufficient coverage of the phase space, so that even small KAM islands can be found.

Overall, the parameter space~$(w, \sigma)$ in Fig.~\ref{fig:tongue}(a) is characterized by a periodic tongue-like structure similar to that in triangular soft Lorentz gas~\cite{PRL}. However, in a square lattice the tongues are significantly wider, resulting from the different geometry including an infinite horizon. Each tongue is characterized by a finite density of quasiballistic trajectories (up to 1.5\% in the figure) -- implying anomalous diffusion. We also encounter parameter areas with zero density for quasiballistic trajectories (white regions) corresponding to normal diffusion. It is interesting that these regions even exist for relatively large interscatterer distances $w$ -- given that the potential is sufficiently soft (large $\sigma$).

In Figs.~\ref{fig:tongue}(b-c), we examine the detailed structure of the tongues at high and low densities of the quasiballistic trajectories, respectively. The zoomed figures show small-scale fringes, even though the amount of noise is relatively high. It is expected that fractal-like tongue structures, such as these examples, are ubiquitous in the parameter space. Unfortunately, due to a finite search grid and ensemble sizes, detailed characterization is numerically tedious.

To further examine the local structure of the faint tongue presented in Fig.~\ref{fig:tongue}(c), a cross-section as a function of~$w$ is presented in Fig.~\ref{fig:tongue}(d). 
Despite increasing the ensemble size to~$\num[print-unity-mantissa=false]{e6}$ in both of these Figs., the curve is relatively noisy.
Hence, no clear conclusions related, for instance, to the scaling properties of the curve in Fig.~\ref{fig:tongue}(d) can be drawn. Nonetheless, the presence of three adjacent areas of anomalous diffusion (centered at $w\sim \num{0.610}$, $w\sim \num{0.611}$, and $w\sim \num{0.613}$) is prominent.

\subsection{Particle displacement distributions}

For the hard-wall infinite horizon Lorentz gas, it has been shown~\cite{bleher1992} that the distribution of the displacement vector~$\vect{r}(t)$, scaled with $ \sqrt{t\ln t}$, approaches a normal distribution. However, this limit is numerically unattainable because of weak convergence.~\cite{bleher1992,szasz2007,Dmitrii2009}.
This result differs from the finite horizon Lorentz gas, where the required scaling factor is proportional to the square root of time ($\propto \sqrt{t}$). Recently, analytical results based on a Lévy walk formalism have been presented for the infinite-horizon square Lorentz gas~\cite{zarfaty,zarfaty2}. This approach converges quickly, making it numerically feasible to verify. In this context, we therefore compare the displacement vector distributions from our soft-wall simulations to the hard-wall results presented by Zarfaty et al.~\cite{zarfaty}.

\begin{figure}[t]
    \includegraphics[width=\columnwidth]{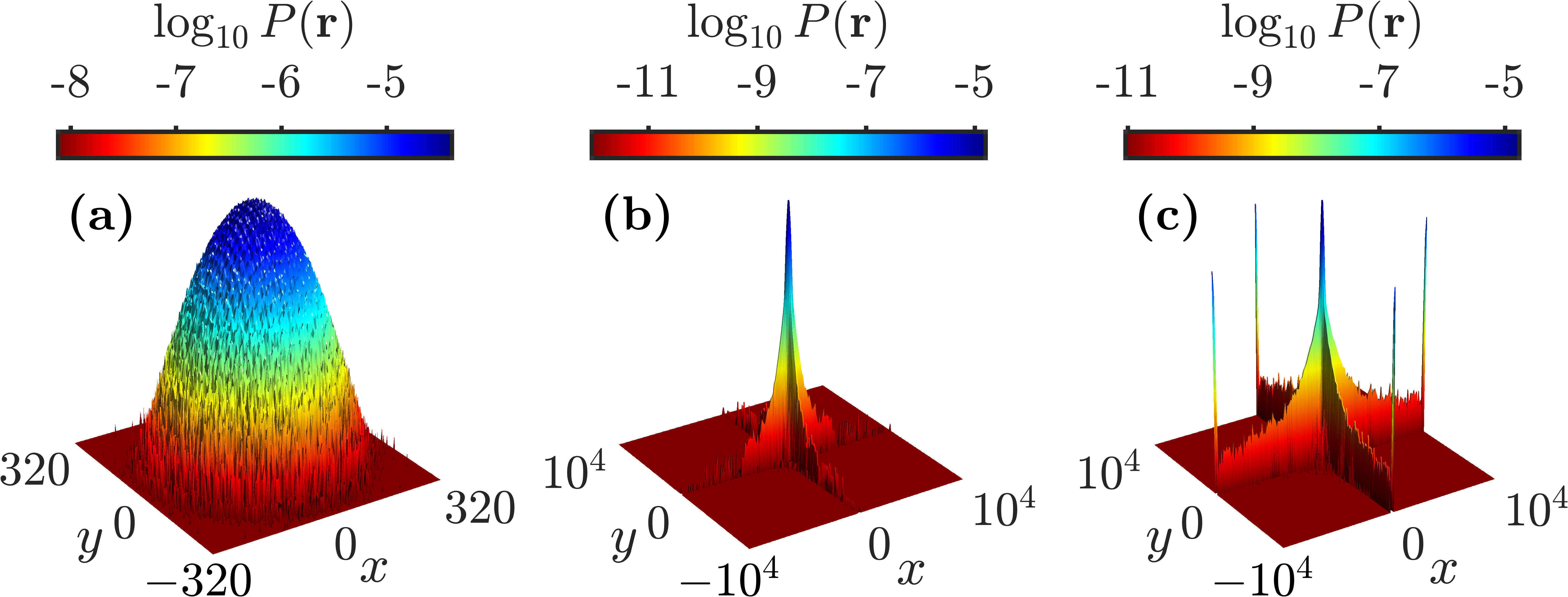}
    \caption{
        Examples of particle displacement distributions in the square soft Lorentz gas at~$t = \num{10000}$ with $w = \num{0.5}$ and varying $\sigma$. The displacements~$x$ and~$y$ are comparable to Ref.~\cite{zarfaty}. (a) Normal diffusion resembling a Gaussian distribution $(\sigma=\num{0.045}\text{, note the logarithmic z-axis})$, (b) Small density of quasiballistic trajectories, leading to long tails along the main coordinate axes $(\sigma\approx\num{0.073})$, (c) High density of quasiballistic trajectories, leading to long tails and peaks at the end $(\sigma=\num{0.1})$.
    }
    \label{fig:dist3d}
\end{figure}

Figure~\ref{fig:dist3d} shows particle distributions at~$t = \num{10000}$. 
We simulated $N = \num[print-unity-mantissa=false]{e6}$ trajectories starting from the center of the unit cell with initial propagation angles between zero and 45 degrees. Owing to the square symmetry, these results cover the entire unit cell through rotation and mirroring. We have selected the parameter values of $\sigma$ and $w$ to enable a direct comparison with Ref.~\cite{zarfaty}.

In Fig.~\ref{fig:dist3d}(a), a typical distribution from system parameters exhibiting normal diffusion is shown, resembling the infinite-time Gaussian limit proven by Bleher~\cite{bleher1992} for the hard-wall Lorentz gas.
Figure~\ref{fig:dist3d}(b) shows a system with a low but non-zero density of quasiballistic trajectories. The distribution is similar to those analytically computed by Zarfaty et al~\cite{zarfaty}, but the tails of our distribution diminish quickly, likely due to the lower trajectory count compared to Ref.~\cite{zarfaty}. Finally, Fig.~\ref{fig:dist3d}(c) depicts a distribution from a system configuration with a high density of quasiballistic trajectories, leading to high probability density along the main coordinate axes, indicative of numerous sticky trajectories or long quasiballistic flights.

\begin{figure}[t!]
    \includegraphics{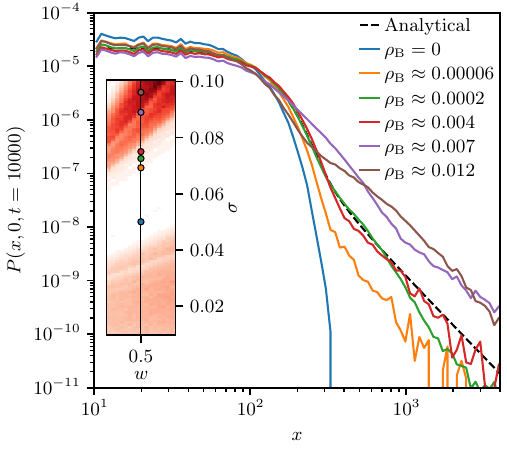}
    \caption{
        Slices of distributions along the main coordinate axis for different densities of quasiballistic trajectories. Similarly to Fig.~\ref{fig:dist3d}, the displacement~$x$ is comparable to Ref.~\cite{zarfaty}.
        The dots in the inset indicate the system parameters in the tongue landscape (Fig.~\ref{fig:tongue}) for each colored line.
    }
    \label{fig:dist}
\end{figure}

On the other hand, Fig.~\ref{fig:dist} compares our results to the analytical distribution at~$t = \num{10000}$ by Zarfaty et al.~\cite{zarfaty}, presenting a \emph{slice} of the distribution along the main coordinate axis.
Again, the system with no quasiballistic trajectories ($\rho_B=\num{0}$) follows a pattern resembling a 2D normal distribution (note the logarithmic y-axis). However, even a slight presence of quasiballistic behavior ($\rho_B\sim\num{0.00006}$) causes deviations from a Gaussian distribution, showing a long tail. As $\rho_\mathrm{B}$ is further increased, the tails become thicker, but the behavior as functions of $x$ and $\rho_B$ is relatively irregular, stemming from both the extreme sensitivity of the system in the regime of anomalous diffusion, and the numerical limitations.
Nevertheless, the analytical limit (dashed line) is well captured with $\rho_B \sim \num{0.004}$.

In general, we find rich dynamics in the diffusion properties of the square soft Lorentz gas.
In parameter areas where normal diffusion is prevalent, the displacement distribution is found to be Gaussian, contrary to the hard-wall gas where the Gaussian distribution is only a theoretical result and unattainable in practice due to enormous computational requirements to reach the long-time limit.
Moreover, in parameter areas where regular motion can be found, we notice that the tails of the distribution are at different heights depending on the volume of the portion of the phase space covered by regular trajectories.

\section{Summary}\label{sec:summary}

In summary, we have  analytically and numerically studied the diffusion properties of a square soft Lorentz gas under various system configurations. The infinite horizon of the system complicates the analysis, and conventional random-walk models require numerical estimates for relevant parameters to accurately describe the behavior. We developed a hopping model that, in the single-hop limit, recovers the MZ approximation and provides reasonable estimates for the numerically computed diffusion coefficient in regimes of normal diffusion as the number of hops increases.

Our findings indicate that diffusion is largely anomalous caused by the presence of quasiballistic orbits, which manifest as long-flight trajectories and appear as KAM islands on the Poincaré surface of section. We have identified a convoluted tongue structure in the density of quasiballistic trajectories, similar to those found in the triangular soft Lorentz gas. Even with the infinite horizon, the square soft Lorentz gas also exhibits localized periodic orbits that either circulate the scatterers or bounce between them in intricate patterns.

Our analysis of the particle displacement distributions reveals notable similarities to analytical and numerical results for a hard-wall square Lorentz gas. Specifically, in the regime of normal diffusion, the distributions exhibited Gaussian behavior, while quasiballistic orbits produced long tails in the distribution. Furthermore, we have pinpointed a softening parameter that yields the analytical result of the hard-wall square Lorentz gas in the long time limit.

An interesting open question is what type of anomalous diffusion is generated whenever there is a quasiballistic orbit, in terms of a known stochastic process (if any) suitably reproducing the original deterministic dynamics in this system. Naively, one might expect a type of L\'evy walk, perhaps as discussed in Refs.~\cite{CGLS14,CGLS15}. Another important open question is to better understand the surprisingly regular tongue structure of quasiballistic orbits reported in Fig.~\ref{fig:tongue}, perhaps along the lines of the mathematical theory developed in Refs.~\cite{turaev1998islands,RKT99}. We hope that this work will not only stimulate further theoretical research on anomalous diffusion in periodic potentials but also applications to experimentally relevant systems such as nanomaterial lattices modeled by realistic soft potentials \cite{StTi20}.

\begin{acknowledgments}
This work was funded by the Research Council of Finland, ManyBody2D Project (No. 349956). The authors acknowledge CSC – IT Center for Science, Finland, for generous computational resources. J.K.-R. thanks the Emil Aaltonen Foundation and the Oskar Huttunen Foundation.
\end{acknowledgments}

\appendix*
\section{Additional details on the numerical simulations}

For the selection of the solver and timestep~$\Delta t$, we performed numerical tests by launching several particles towards one scatterer at different impact parameters~$b$ and measured the average absolute deviation of the total energy from the expected value of~$E = 1/2$ during the deflection.
These tests were repeated for different solvers, timesteps~$\Delta t$ and softness parameteres~$\sigma$.
The solver abbreviations used below are listed in Table~\ref{tab:solvers}.

\begin{table}[ht]
    \centering
    \caption{Symplectic solvers supported by the Bill2D~\cite{bill2d} software.}
    \begin{tabular}{ccc}
         \textbf{Abbreviation} & \textbf{Solver} & \textbf{Ref.} \\\hline
         2                     & Velocity Verlet                 & \cite{compsim} \\
         2opt                  & 2nd order, optimal coefficients & \cite{theintegrator} \\
         3                     & 3rd order                       & \cite{Suzuki1992integrator} \\
         4                     & 4th order                       & \cite{Yoshida1990integrator} \\
         4opt                  & 4th order, optimal coefficients & \cite{theintegrator} \\
         6opt                  & 6th order, optimal coefficients & \cite{theintegrator}
    \end{tabular}
    \label{tab:solvers}
\end{table}

\begin{figure}[th!]
\includegraphics{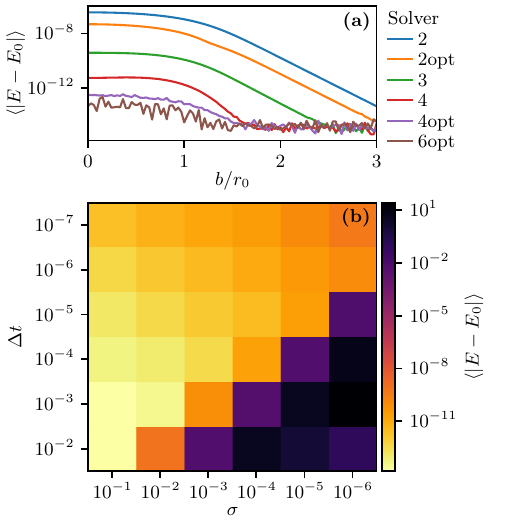}
\caption{
    Average absolute deviation of energy of the Bill2D software~\cite{bill2d}, which measures the accuracy of the computation. (a) Accuracy for different solvers with $\Delta t = \num{e-3}$ and $\sigma = \num{0.05}$. (b) Accuracy as functions of $\Delta t$ and $\sigma$ with the 6opt solver.
}
\label{fig:numacc}
\end{figure}

The results for~$\Delta t = \num{e-3}$ and~$\sigma = \num{0.05}$ are shown in Fig.~\ref{fig:numacc}(a). Among the tested solvers, we find that the 6opt solver yields the lowest average absolute energy deviation. Based on this, we select 6opt for further detailed analysis below.

In Fig.~\ref{fig:numacc}(b) we examine the accuracy as functions of both the timestep $\Delta t$ and the softness parameter $\sigma$. Our findings indicate that as $\sigma$ decreases, a smaller timestep is required to maintain accuracy. However, for a fixed $\sigma$, there exists an optimal $\Delta t$ that provides the best accuracy. In the range of parameters examined, the optimal timestep is typically $0.1\ldots 1$ times $\sigma$.

From these results, we conclude that the optimal solver is a 6th-order method with optimal coefficients~\cite{theintegrator}, and that the timestep must be adapted based on the softness of the potential. Notably, reducing the timestep significantly increases the computational cost.

The challenge of developing an adaptive timestep solver remains an open question for future work. Such a method would be advantageous for simulating systems with small $\sigma$, where particle trajectories alternate between (nearly) linear motion in open regions and more complex behavior near scatterers.



\providecommand{\noopsort}[1]{}\providecommand{\singleletter}[1]{#1}%

\end{document}